# Molecular beam epitaxy of the magnetic Kagome metal FeSn on LaAlO$_3$ (111)

Deshun Hong[1], Changjiang Liu[1], Haw-Wen Hsiao[2], Dafei Jin[3], John E. Pearson[1], Jian-Min Zuo[2] and Anand Bhattacharya[1,#]

[1] Materials Science Division, Argonne National Laboratory, Argonne, Illinois 60439, USA;

[2] Department of Materials Science and Engineering, University of Illinois at Urbana-Champaign, Urbana, Illinois 61801, USA;

[3] Center for Nanoscale Materials, Argonne National Laboratory, Argonne, Illinois 60439, USA;

**Abstract:** Materials with Kagome layers are expected to give rise to rich physics arising from band structures with topological properties, spin liquid behavior and the formation of Skyrmions. Until now, most work on Kagome materials has been performed on bulk samples due to difficulties in thin film synthesis. Here, by using molecular beam epitaxy, layered Kagome-structured FeSn films are synthesized on (111) oriented LaAlO$_3$ substrate. Both in-situ and ex-situ characterizations indicate these films are highly crystalline and *c*-axis oriented, with atomically smooth surfaces. The films grow as disconnected islands, with lateral dimensions on the micron meter scale. By patterning Pt electrodes using a focused electron beam, longitudinal and transverse resistance of single islands have been measured in magnetic fields. Our work opens a pathway for exploring mesoscale transport properties in thin films of Kagome materials and related devices.

[#]Email: anand@anl.gov

Topological properties of a material can be mapped to the particular symmetry observed in a class of material systems. This is true for topological semimetals, such as Dirac semimetal or Weyl semimetal. For example, the Dirac nodes in $Cd_3As_2$ are protected by a $C_4$ rotation symmetry [1,2,3], while the existence of Weyl nodes in the Kagome material $Mn_3X$ (X = Ge [4, 5], Sn [6, 7]) are protected by the glide mirror symmetries, considering both magnetic and crystal structures. Furthermore, there is significant interest in understanding the interplay between electronic correlations and topology, potentially giving rise to collective states. So far, however, most works related to topological semimetals have been focused on bulk crystals, which hinder the tunability of these topological properties, as well as applications in devices. Furthermore, atomically flat surfaces are hard to achieve by cleavage in certain materials, which makes probing of band structure by angular resolved photoemission spectroscopy (ARPES) or scanning tunneling microscopy (STM) challenging. Therefore, thin film synthesis of these topological materials is of great importance for both fundamental research and applications.

Until now, there have only been a few materials studied in thin films such as Bi-Sb alloys [8], $Bi_2Se_3$ [9], $Bi_2Te_3$ [10] and honeycomb structured two dimensional materials as $WTe_2$ [11], $WSe_2$ [12], with the layered materials attracting additional interest due to the emergence of "twistronics", building upon recent works on graphene[13, 14]. Distinct from the honeycomb lattice in graphene, the hexagonal Kagome structure is composed of corner sharing triangles, which also hosts rich physics like linear band dispersion and flat band[15], spin liquid behaviors [16-18] and large spin Hall angle in non-collinear antiferromagnets as $Mn_3X$ (X = Ir [19], Ge [4, 5], Sn [6, 7]).

The Kagome semimetal $Fe_3Sn_2$ hosts Skyrmions [20] in a wide temperature range, and has drawn attention due to its rich electronic properties. As a ferromagnet below room temperature, each $Fe_3Sn_2$ unit cell is composed of multiple layers (six $Fe_3Sn$ layers and three $Sn_2$ layer). The $Fe_3Sn$ layers are Kagome structured with Fe atoms forming corner sharing triangles with Sn atoms sitting in the center of Fe hexagons. Very recent experimental works on $Fe_3Sn_2$ show that the electronic structure shows both flat bands and linearly dispersion features near the Fermi level [21]. By applying an external magnetic field, the electronic structure of state can be tuned and a nematic state was revealed, suggesting the important role of correlations [22]. Theories also suggest $Fe_3Sn_2$ may be a candidate for high temperature quantum anomalous hall effect [23]. Beside Kagome structured $Fe_3Sn$ layer, a single layer of stanene (honeycomb $Sn_2$) is interesting as it is supposed to be topological nontrivial in the 2D limit [24]. By playing with stacking sequence of $Fe_3Sn$ and $Sn_2$, we can get several interesting $(Fe_3Sn)_xSn_{2y}$ compound. Among them, FeSn is composed of one layer of Kagome $Fe_3Sn$ and one layer of stanene $Sn_2$. ARPES experiment have revealed that FeSn hosts Dirac fermions and flat bands very recently [25], which suggest $(Fe_3Sn)_xSn_{2y}$ a big playground for artificial topological materials by varying stacking order.

Previous experiments on both $Fe_3Sn_2$ and FeSn were performed on cleaved bulk materials. Here, by using molecular beam epitaxy, we have synthesized FeSn films on (111) oriented $LaAlO_3$ substrate. Both in-situ and ex-situ characterization results indicate the high crystalline of our film with flat surface. By using focused electron beam epitaxy, we patterned nano-electrodes on a single island using electron-beam assisted deposition and characterized both longitudinal and transverse resistance as a function of temperature and magnetic field.

Fig. 1 shows the lattice structure of FeSn. It is hexagonal and has the same structure as CoSn, belonging to the space group of P 6/m m m (191). The lattice constants *a*, *b* and *c* of FeSn are 5.297 Å, 5.297 Å and 4.481 Å, respectively. Each unit cell is composed by two layers: $Fe_3Sn$ and $Sn_2$. The atomic arrangement in these layers can be more clearly seen in Fig. 1 (b) and Fig. 1 (c). For the Kagome layer (Fig. 1 (b)), Fe atoms form corner sharing triangles while Sn atoms fill in the center of each Fe hexagons. For the stanene layer (Fig. 1 (c)), Sn atoms arrange similarly as carbon atoms in graphene. Although non-buckled stanene films are tricky to be synthesized directly [25, 26], $Fe_3Sn$ layer can stabilize this 2-dimensional honeycomb easily when $Sn_2$ layers are sandwiched in between, making it a new playground for proximity effect related topological physics.

FeSn films are synthesized using molecular beam epitaxy (MBE). Considering its crystal structure and lattice constant, we chose (111) oriented LaAlO$_3$ (LAO) as substrate, which can be viewed as hexagonal with lattice constant of 5.36 Å (~ 1% mismatch with FeSn). Fe and Sn sources are evaporated in separate effusion cells at 1305 °C and 1120 °C correspondingly, and the growth rates are measured by quartz crystal microbalance (QCM) and is further calibrated with Rutherford backscattering spectrometry (RBS). Fe and Sn flux are $1.94 \times 10^{13} cm^{-2} \cdot s^{-1}$ and $1.52 \times 10^{13} cm^{-2} \cdot s^{-1}$ during the growth. The stoichiometry of our film is precisely controlled by the shutter time of Fe and Sn during co-deposition. Before growth, LAO substrates are sonicated in acetone and isopropyl alcohol for 5 minutes each, and then degassed in vacuum in a preparation chamber at 600 °C until the pressure is lower than 5 x 10$^{-9}$ Torr. Then the substrate is transferred to the growth chamber and further annealed at 900 °C to improve crystal quality. We use 500 °C for growth and the growth process is monitored by reflection high energy electron diffraction (RHEED) with an acceleration voltage of 10 kV.

As can be seen in Fig. 2 (a) and Fig. 2 (b), sharp RHEED streaks viewing along $[1\bar{1}0]$ and $[11\bar{2}]$ indicate the high crystal quality and flat surface of the film after growth. By comparing the spacing of the Laue dots before (substrate) and after growth (film), we find that the film is not strained to the LaAlO$_3$ substrate but has the same a and b lattice constant as bulk. We carried out X-ray diffraction measurements using Cu K$_{\alpha 1}$ radiation, as shown in Fig. 2 (c). The film is mostly c axis oriented with an out-of-plane lattice constant 4.45 Å, except a small amount of another orientation (peak near 44.5°) which is (012) plane. We further measured the morphology of the films, as can be seen in Fig. 2 (d) and Fig. 2 (e). The films are discontinuous and composed of micron sized islands. This discontinuity may be related to the high surface energy of the film. These islands are much larger than the coherence length of our RHEED beam, though smaller than the footprint (100's of microns) – the sharp RHEED pattern indicates on average a large fraction of epitaxially oriented islands. According to our AFM result, some of these islands are hexagonally shaped, reflecting the hexagonal crystal structure. The islands are atomically flat and terraces with step size of ~ 4.5 Å can be seen when scanning on more localized area. In between these big islands, there are small clusters which may produce the (012) peak in the XRD. We note that our approach differs from that in similar work by H. Inoue et. al. [26] where they use $SrTiO_3$(111) which has bigger lattice constant. Further more, our films are directly grown in a crystalline structure at high temperatures, without a capping layer, and do not require post growth annealing. The atomically smooth exposed surfaces of our films are favorable to characterizations by probes such as ARPES and STM for direct mapping of band structures.

FeSn films were further characterized by scanning transmission electron microscope (STEM). The sample was prepared by focused ion beam (FIB). To protect the film, a thin layer of carbon was deposited on top, and the film is stable during processing. As can be clearly seen in Fig. 3 (a), the ~ 35 nm thick film is homogenous with a flat top surface. However, for the interface between film and LAO substrate, there is a grey region showing different contrast with the film on top. This blurred area is about 5 nm thick, which is consistent with RHEED pattern evolution during the early stages of growth. At the beginning of the FeSn growth, the RHEED pattern blurs quickly and re-appears after ~ 10 unit cells (after which the RHEED streaks become sharper). We zoomed in the top surface of the film and show a high-resolution image in Fig. 3 (b). As can be seen, two types of layers are stacking alternatively. Further analysis shows that these two layers are $Fe_3Sn$ layer and $Sn_2$, and atomic structure is the same as FeSn when viewing from $[10\bar{1}0]$ direction.

Structural analysis presented above indicates that our film is composed of c-axis oriented FeSn islands. The electronic transport measurements obtained on these islands are shown in Fig. 4. Since our film is discontinuous, direct resistance measurement on the whole film shows an open circuit. Therefore, we patterned ~ 200 nm wide Pt electrodes on single island by using electron beam assisted deposition in a scanning electron microscope (SEM). An image of the device is shown in the inset of Fig. 4 (a). The temperature dependence of resistance measured in two separate devices show similar metallic behavior, with residual resistance ratio (RRR) 15 and 6.4, respectively.

one order lower than that measured on bulk [27], presumably due to defects and scattering from surfaces. Field dependence of both $R_{xy}$ and $R_{xx}$ at various temperatures in magnetic fields applied perpendicular to the plane are further measured, as shown in Fig. 4(c) and Fig. 4(d). At temperatures of about 250 K a clear nonlinearity shows up in $R_{xy}$, similar to the recent results on FeSn film grown on $SrTiO_3$ [26], which is ascribed to existence of multi-bands from Sn network. In Fig. 4 (d), small negative magnetoresistance is observed at high temperatures while positive magnetoresistance (MR) begins to show up below 50 K. The amplitude of MR increases upon cooling. The positive magnetoresistance shows a parabolic shape, consistent with orbital origin of the magnetoresistance.

In conclusion, by using high vacuum MBE, we grew FeSn films on (111) oriented $LaAlO_3$. Both in-situ RHEED and ex-situ XRD and AFM results indicate that the bulk of the FeSn film is highly crystalline despite having an initial amorphous layer on the substrate. STEM results show a highly ordered alternating layered structure of $Fe_3Sn$ and $Sn_2$. By patterning nano-electrodes on single FeSn island, electronic transport properties are measured in various temperatures and magnetic fields, consistent with those found in bulk single crystals. Further works are needed as solving non-wetting issue as well as reaching 2D limit for high temperature quantum anomalous Hall effect. Our work opens up the possibility of exploring mesoscale transport and surface properties of the rich electronic states found in Kagome magnetic metals.

Work at Argonne was supported by the Center for the Advancement of Topological Semimetals, an Energy Frontier Research Center funded by the U.S. Department of Energy (DOE), Office of Basic Energy Sciences. The use of facilities at the Center for Nanoscale Materials was supported by the U.S. DOE, Office of Basic Energy Sciences under Contract No. DE-AC02-06CH11357.

**Data Availability Statement:** The data that support the findings of this study are available from the corresponding author upon reasonable request.

**Fig. 1**

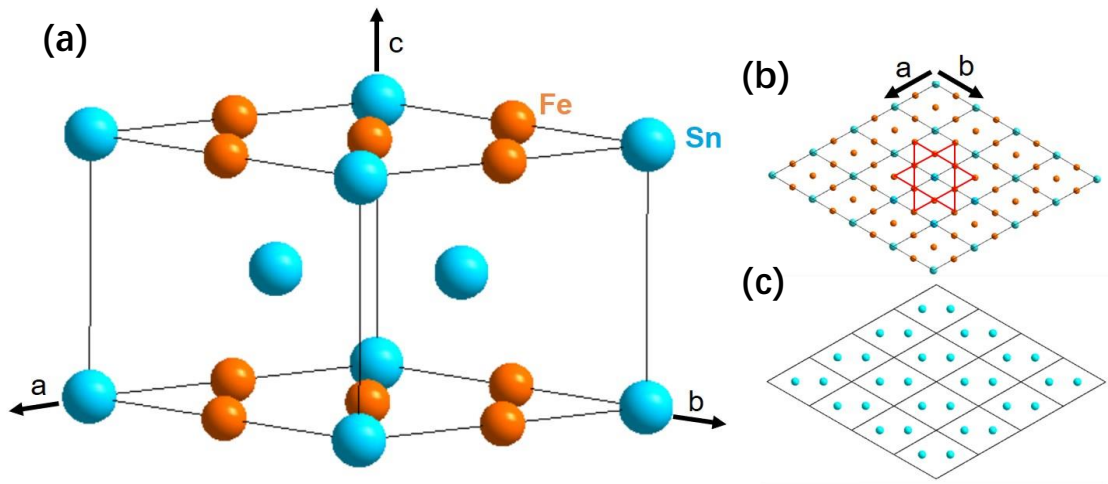

Fig. 1 (a) Crystal structure of FeSn which is composed of Kagome $Fe_3Sn$ and $Sn_2$ layers. 4 x 4 $Fe_3Sn$ (b) and $Sn_2$ layers (c). The corner sharing Fe triangles with Sn in the center is outlined in (b).

**Fig. 2**

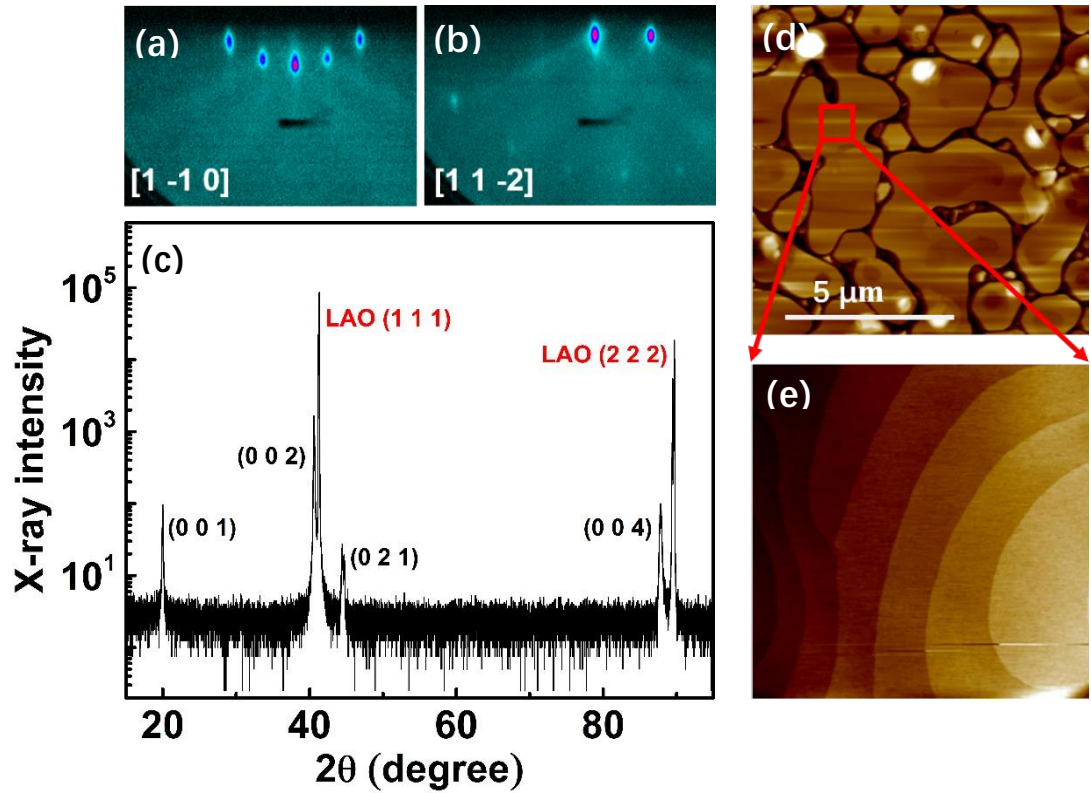

Fig. 2 (a) and (b) RHEED image viewing along [1 -1 0] and [1 1 -2] after growth. (c) X – ray diffraction of FeSn film using Cu-$K_{\alpha 1}$. (d) AFM image of FeSn film after growth and the film shows discontinuity. (e) Zoom – in scan (1 um x 1 um) on one island from (d) indicates the atomically flat surface and terraces.

**Fig. 3**

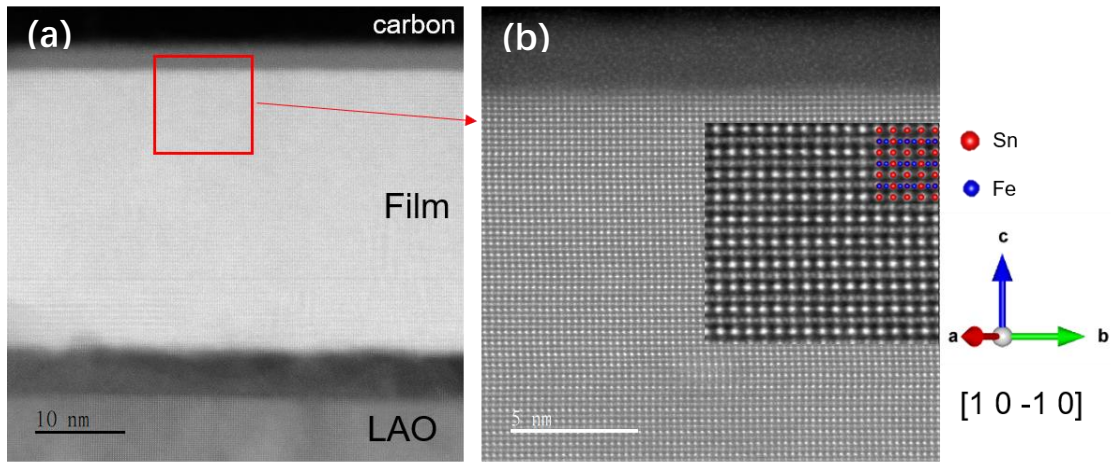

Fig. 3 (a) STEM image of cross section of FeSn film, with carbon as protecting layer. (b) Zoomed in image from (a). Inset of (b): layered structure can be clearly seen and atomic structure is the same as theoretical model when viewing along [1 0 -1 0] direction.

**Fig. 4**

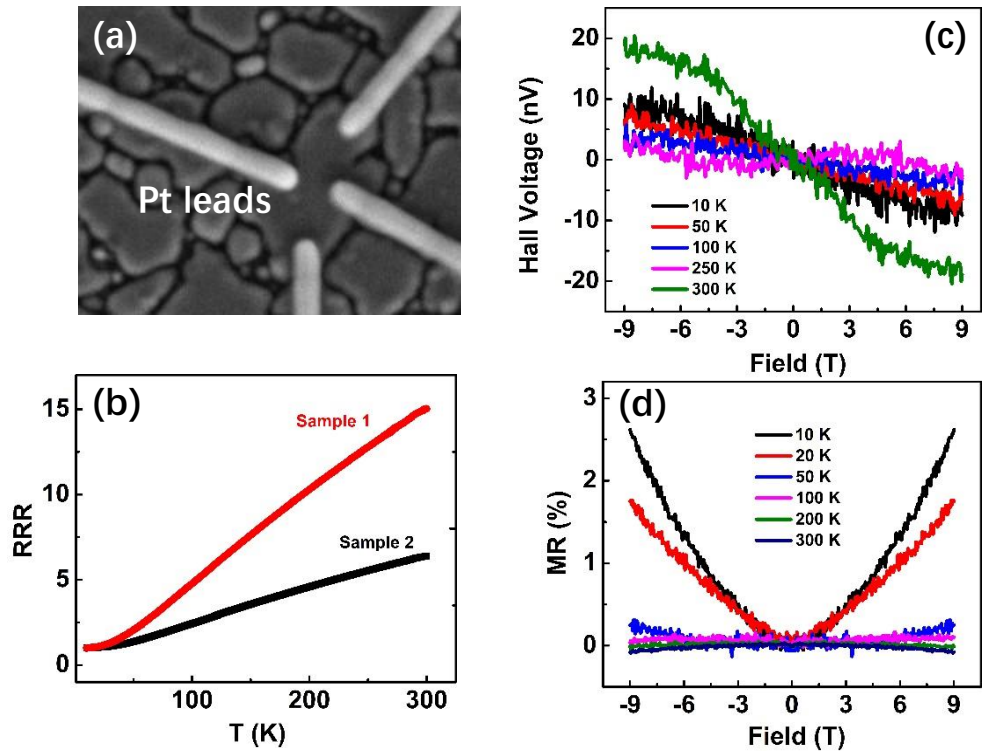

Fig. 4 (a) SEM image of patterned device with for Pt leads on single island. Width of Pt wires is ~ 200 nm. (b) Temperature dependence of RRR in two devices, and both devices show metallic behaviors. (c) Hall signals measured at various temperatures with positive slopes indicating electron dominated carriers ($I_{bias}$ = 1 uA). Anomaly begins showing up at above 250 K. (d) Magnetoresistance at various temperature showing parabola behavior.